\documentclass[10pt,prb,twocolumn,showpacs]{revtex4}

\usepackage{amsmath,amssymb} 
\usepackage{graphicx}        
\usepackage{hyperref}        

\begin{document}
\title{Effect of Disorder on the Superfluid Transition in Two-Dimensional Systems}%
\author{Kar\'en G. Balabanyan}%
\affiliation{Department of Physics, University of Massachusetts, Amherst, Massachusetts 01003, USA}%
\date{\today}%
%
%
%
%
\begin{abstract}
In recent experiments on thin $^4$He films absorbed to rough surfaces Luhman and Hallock (Ref.~\onlinecite{b:exp:SFN:FiniteSize:Dwight}) attempted to observe KT features of the superfluid--normal transition of this strongly disordered 2D bosonic system. It came as a surprise that while peak of dissipation was measured for a wide range of surface roughness there were no indications of the theoretically expected universal jump of the areal superfluid density for the strongly disordered samples. We test the hypothesis that this unusual behavior is a manifestation of finite-size effects by numerical study of the corresponding 2D bosonic model with strong diagonal disorder. We demonstrate that the discontinuous features of the underlying KT transition are severely smoothed out for finite system sizes (or finite frequency measurements). We resolve the universal discontinuity of the areal superfluid density by fitting our data to the KT renormalization group equations for finite systems. In analogy to our simulations, we suggest that in experiments on strongly disordered 2D bosonic systems the very existence of the KT scenario can and should be revealed only from a proper finite-size scaling of the data (for $^4$He films finite-size scaling can be effectively controlled by the scaling of finite frequency of measurements). We also show relevance of our conclusions for a wider class of systems, such as superconducting granular films, Josephson junction arrays, and ultracold atomic gases, where similar difficulties appear in experiments designed to verify KT transition (especially in disordered cases).
\end{abstract}
%
%
%
%
\maketitle
%
%
%
%
%
%
\section{Introduction}
%
%
%
%
For decades $^4$He films played a prominent role in the research of superfluidity of two-dimensional (2D) bosonic systems. The Kosterlitz-Thouless (KT) theory\cite{b:KT:1} describes the finite temperature superfluid phase transition for the $^4$He films absorbed to the flat substrates.\cite{b:exp:SFN:He1} One of its main predictions is the universal jump from zero of the areal superfluid density $\sigma_\mathrm{SF}$ at the transition temperature $T_\mathrm{cr}$ given by the relation\cite{b:KT:UnivJump77}
\begin{equation}
\frac{2}{\pi}=\frac{\hbar^2}{m^2k_\mathrm{B}T_{\mathrm{cr}}}\cdot{\sigma_\mathrm{SF}}_\mathrm{cr}%
,\label{eqn:FT:rhoKT}%
\end{equation}
where $m$ is the mass of a $^4$He atom. Theoretically, according to Harris criterion, disorder is irrelevant at the KT transition, and should not alter the discontinuous KT scenario.\cite{f0} Recently, Luhman and Hallock (Ref.~\onlinecite{b:exp:SFN:FiniteSize:Dwight}) performed extensive quartz crystal microbalance (QCM) experiments of the superfluid transition of the $^4$He films absorbed to rough surfaces. Surprisingly, their observations were not directly agreeing with the expected KT scenario of the 2D superfluid transition. They found that it was increasingly difficult to identify signatures of the universal jump (\ref{eqn:FT:rhoKT}) as the surface roughness was increased (see also Ref.~\onlinecite{b:exp:SFN:FiniteSize:Dwight3dSound}). For strong enough disorder the presence of the KT transition was seen only indirectly by the observation of a peak in the dissipation. Could their observations be an indication of some strong-randomness scenario of the phase transition, or is there a way to reconcile those experiments with the standard KT theory?

Note that for ideal (disorder free) 2D bosonic systems discontinuity of the superfluid density at the critical point is a characteristic of the phase transition in the thermodynamical limit of infinite system sizes. In the experiments with the superfluid helium films the sample size can simply set finite length scales, e.g., it can be the size of a confining channel,\cite{b:exp:SFN:FiniteSize:WilliamsBoundaries} or the radius of spherical grains $R$ in the case of helium films absorbed in packed powders,\cite{b:exp:SFN:FiniteSize:Williams} or the pore radius $R$ in the case of helium films absorbed on porous glass\cite{b:exp:SFN:FiniteSize:PorousGlass} (in the latter two cases crossover from the effective 2D to 3D behavior should be kept in mind\cite{f1}). Finite length scales can also be related to finite frequency of measurements, when one is bounded by the frequency-dependent vortex diffusion length $r_\omega=\sqrt{14D/\omega}$, where $D$ is the vortex diffusivity and $\omega$ is the experimental frequency.\cite{b:KT:AHNS,b:KT:Donnelly} Finite-size effects in experiments smear the universal jump (\ref{eqn:FT:rhoKT}), and disorder is expected to increase this finite-size broadening of the KT transition even more. We think this scenario is applicable to the finite frequency measurements on $^4$He films absorbed to the rough surfaces in Ref.~\onlinecite{b:exp:SFN:FiniteSize:Dwight}. The purpose of this study is to show that in these experiments disorder of the substrate \emph{can} produce such dramatic additional broadening of the superfluid transition on top of finite-size effects so that it could be indeed difficult to resolve the growth of the superfluid fraction on the varying normal fluid background.

In Sec.~\ref{s:FT:Simulations} we start with the description of $^4$He films absorbed to rough surfaces studied in Ref.~\onlinecite{b:exp:SFN:FiniteSize:Dwight} by a 2D bosonic model with diagonal disorder. We want to be as close in our simulations to the experimental situation of Ref.~\onlinecite{b:exp:SFN:FiniteSize:Dwight} as possible in order to guarantee that all physics that might be important for explanation of surprising outcome of Ref.~\onlinecite{b:exp:SFN:FiniteSize:Dwight} will be preserved. We represent the microscopically disordered rough surface by a random potential (i.e., diagonal disorder). We keep temperature fixed, and tune the superfluid transition by adjusting the chemical potential (the average helium film thickness is a monotonic increasing function of the chemical potential). In the ideal case, we observe the anticipated finite-size broadening of the universal jump of the superfluid density (\ref{eqn:FT:rhoKT}) near the critical point of the KT transition. Inclusion of the strong disorder leads to such severe additional broadening that, compared to the ideal case, it may seem impossible to make any definite conclusions about the underlying transition type. Our analysis utilizes Monte Carlo simulations of different system sizes, or, equivalently, ability to set $r_\omega$ by carrying out measurements at different finite frequencies. We resolve the asymptotic universal jump of the superfluid density (\ref{eqn:FT:rhoKT}) by showing that data do obey the KT renormalization group (RG) equation. Finally, once we have established the KT scenario in the strongly disordered case, in Sec.~\ref{s:FT:Comparison} we carry out a qualitative comparison of our simulations with the aforementioned Ref.~\onlinecite{b:exp:SFN:FiniteSize:Dwight} QCM experiments with $^4$He films absorbed to rough surfaces. In the context of the QCM technique, we also discuss how the dissipation peak is affected by disorder.

With respect to the KT transition, we expect the combined effect of diagonal disorder and finite-size broadening to be qualitatively similar for different physical systems, and to be independent of the particular type of disorder realization and choice of a control parameter. In Sec.~\ref{s:FT:Related} we discuss how our results for the $^4$He films can be related to superconducting granular films, Josephson junction arrays, and ultracold atomic gases.
%
%
\section{Simulations\label{s:FT:Simulations}}
\begin{figure}
\centering
\includegraphics[width=0.65\columnwidth,keepaspectratio]{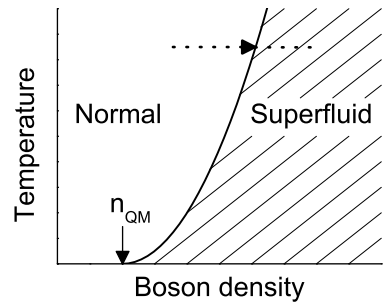}
\caption{\label{fig:FT:phaseDIAG}Generic phase diagram in the density-temperature plane for 2D bosonic system with strong diagonal disorder ($n_{\mathrm{QM}}$ is the threshold boson density). We study the superfluid transition along the dotted line path.}
\end{figure}

%
%
In Fig.~\ref{fig:FT:phaseDIAG} we show a phase diagram in the temperature-density plane for a 2D bosonic model (\ref{eqn:FT:Hubbard}) with strong diagonal disorder. The transition from the normal to superfluid phase happens according to the KT scenario at the phase boundary. At some finite temperature $T$ we drive the system through the transition by increasing the boson density. At the critical point, the superfluid areal density experiences the universal jump ${\sigma_\mathrm{SF}}_\mathrm{cr}$ according to~(\ref{eqn:FT:rhoKT}).

For a system of a given finite size, the superfluid density is expected to be a monotonic function of the boson density, and we study its sharpness with respect to disorder. The strength of disorder also determines the position of the phase boundary.

Note that the zero temperature quantum phase transition is in a different universality class.\cite{b:SFI:FWGF89} In the case of $^4$He films on disordered substrates, the threshold boson density $n_\mathrm{QM}\lesssim 1$ per lattice site corresponds to up to several atomic layers.\cite{b:SFI:exp:Reppy1PRB}

%
%
\subsection{Model\label{s:FT:Simulations:model}}
%
%
We start with the description of bosonic systems with diagonal disorder by a general lattice bosonic Hubbard model:%
\begin{equation}\label{eqn:FT:Hubbard}
H= -t\sum_{\left\langle i,j\right\rangle}%
   \left(\hat{a}_i^\dag\hat{a}_j+\hat{a}_i\hat{a}_j^\dag\right)%
   +\frac U2\sum_i\hat{n}_i^2%
   -\sum_i\left( \mu + v_i\right)\hat n_i,%
\end{equation}
where $\hat{a}_i^\dag$ and $\hat{a}_i$ are the creation and annihilation operators, $\hat{n}_i=\hat{a}_i^\dag\hat{a}_i$ is the number operator on site $i$, and $\left\langle i,j\right\rangle$ denotes summation over pairs of nearest neighbors, each pair counted once. The first term describes nearest neighbor hopping, and the second term corresponds to soft-core on-site repulsion. The chemical potential $\mu$ defines the average number of bosons per site $n(\mu)$, whereas $v_i$ represents random on-site potential. The value of $v_i$ is distributed uniformly between $-\Delta$ and $\Delta$.

We do not carry out quantum Monte Carlo simulations of the Hubbard model (\ref{eqn:FT:Hubbard}) in continuous time.\cite{b:WA:ProSvi0}. Instead, we map 2D quantum bosonic model (\ref{eqn:FT:Hubbard}) on the equivalent (2+1)D classical $J$-current model.\cite{b:WAsim:MAIN:WSGY} For this classical (2+1)D counterpart of quantum 2D model (\ref{eqn:FT:Hubbard}) we perform classical Monte Carlo simulations with high-performance worm algorithm.\cite{b:WA:ProSvi1,b:WAsim:ProSviLONG} Action of the so-called $J$-current model is written in terms of integer link-current variables $\mathbf{J}=(J^x,J^y,J^\tau)$ defined on a $N\times N\times N_\tau$ space and imaginary time lattice ($N_\tau\delta\tau=\hbar/k_\mathrm{B}T$, where $\delta\tau$ is an imaginary time interval):
\begin{equation}
S=\frac{\tilde{t}}{2}\sum_\mathbf{n}%
\left[\big{(}J_\mathbf{n}^x\big{)}^2+\big{(}J_\mathbf{n}^y\big{)}^2\right]
+\frac{\tilde{U}}{2}\sum_\mathbf{n}%
\left[\big{(}J_\mathbf{n}^\tau\big{)}^2-%
(\tilde{\mu}+\tilde{v}_\mathbf{r})J_\mathbf{n}^\tau\right]%
,\label{eqn:FT:H_bond}
\end{equation}
where $\mathbf{n}=(\mathbf{r},\tau)$ are discrete space and imaginary time coordinates. $\mathbf{J}$ has to be conserved and therefore satisfies the local zero-divergence constraint, $\mathbf{\nabla}\cdot\mathbf{J}_\mathbf{n}=0$. Spatial currents $J^x$ and $J^y$ represent hopping, and imaginary time current component $J^\tau$ represents occupation number. We choose to consider only positive occupation numbers, $J^\tau\geq 0$. This is a natural choice for atomic systems, and it cannot affect the universal properties of the transition.\cite{b:WAsim:ProSviLONG} Correspondence of parameters of the models (\ref{eqn:FT:Hubbard}) and (\ref{eqn:FT:H_bond}) are given by
\begin{equation}
\tilde{t}=-2\ln (n_0t\cdot\frac{\delta\tau}{\hbar}),\quad\tilde{U}=U\cdot\frac{\delta\tau}{\hbar},%
\quad\tilde{\mu}=\frac{\mu}{U},%
\quad\tilde{v}_\mathbf{r}=\frac{v_\mathbf{r}}{U}%
,\label{eqn:FT:params}
\end{equation}
where $n_0$ is some offset boson density of model (\ref{eqn:FT:Hubbard}).\cite{f2} Here the value of $\tilde{v}_\mathbf{r}$ is uniformly distributed between $-\tilde{\Delta}$ and $\tilde{\Delta}$, where $\tilde{\Delta}=\Delta/U$.

The total boson density $\sigma$ (proportional to the boson number per site $n$) and the superfluid density $\sigma_\mathrm{SF}$ are determined from Monte Carlo estimators of the winding numbers $W_i=1/N_i\sum_{\mathbf{n}}J_\mathbf{n}^i$, and are given by\cite{b:WAsim:MAIN:WSGY,b:WA:CeperleyWindingNumber}
\begin{subequations}
\label{allequations}
\begin{eqnarray}
\sigma=\frac{\tilde{m}}{a^2}\cdot n%
&,&n=\frac{\left\langle W_\tau\right\rangle}{N^2}%
,\label{eqn:FT:n}\\
\sigma_\mathrm{SF}=\frac{\tilde{m}^2k_\mathrm{B}T}{\hbar^2}\cdot K%
&,&K=\frac{\left\langle W_x^2\right\rangle+\left\langle W_y^2\right\rangle}{2}%
,\label{eqn:FT:rho}
\end{eqnarray}
\end{subequations}
where $\tilde{m}=\hbar^2/2ta^2$ is an effective boson mass on a lattice with spacing $a$. In 2D, the statistics of spatial winding numbers are essentially discrete, thus an appropriate way to find $K$ is through fitting collected winding number histograms with a discrete gaussian $P(W_x,W_y)\propto\exp(-(W_x^2+W_y^2)/2K)$.

For definiteness, we choose numerical values of parameters to be close to those characteristic for experiments with $^4$He films at temperatures of about $1\:\mathrm{K}$, i.e., hopping, interaction and temperature are considered to be of the same order $t\sim U\sim k_\mathrm{B}T$, and $n_0\sim n_\mathrm{QM}\sim 1$ to correlate with a transition of several layers of a $^4$He film. Corresponding effective parameters of the (2+1)D classical model (\ref{eqn:FT:H_bond}) are $\tilde{t}$, and $\tilde{U}$. Our choice of finite $L_\tau$ is a compromise between the efficiency of discrete imaginary time codes and staying very close to the original model. We have varied $N_\tau$, while $\tilde{t}$ and $\tilde{U}$ can be estimated from (\ref{eqn:FT:params}), so that the boson number density in the vicinity of the superfluid transition in the ideal system would stay about 1 per lattice site. We found $\tilde{t}=4$, $\tilde{U}=0.2$ and $N_\tau =6$ to be appropriate values.

\begin{figure}
\centering
\includegraphics[width=\columnwidth,keepaspectratio]{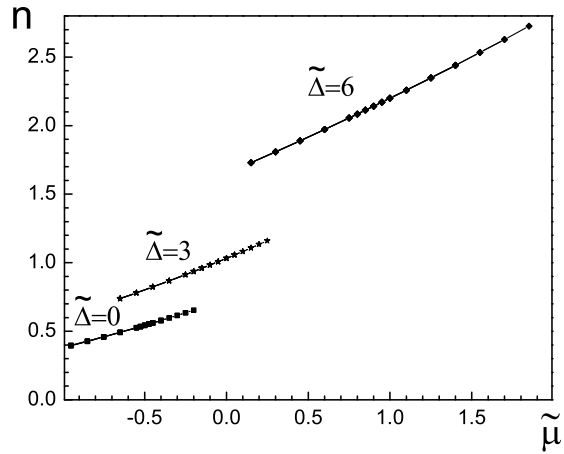}
\caption{\label{fig:FT:n}Average number of bosons per lattice site $n$  as a function of chemical potential $\tilde{\mu}$. Three sets of curves represent the cases of no disorder $\tilde{\Delta}=0$, and disorder strengths $\tilde{\Delta}=3$ and $\tilde{\Delta}=6$. Statistical errors are smaller than the symbol size. Data in each set for different system sizes $24\times 24$, $48\times 48$, $96\times 96$ and $192\times 192$ collapse on one and the same curve, i.e., $n$ is independent of system size.}
\end{figure}
%
%
\subsection{Results\label{s:FT:Simulations:results}}
\begin{figure}
\centering
\includegraphics[width=\columnwidth,keepaspectratio]{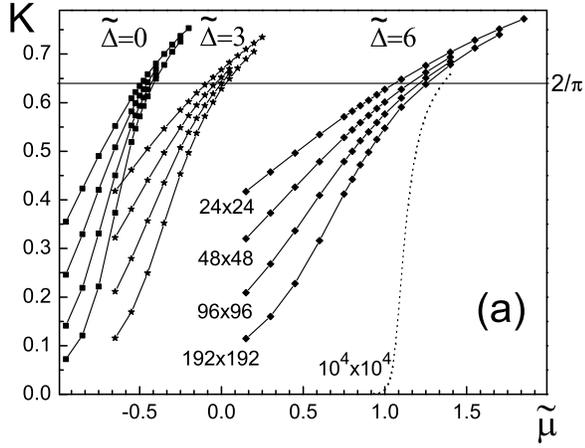}
\includegraphics[width=\columnwidth,keepaspectratio]{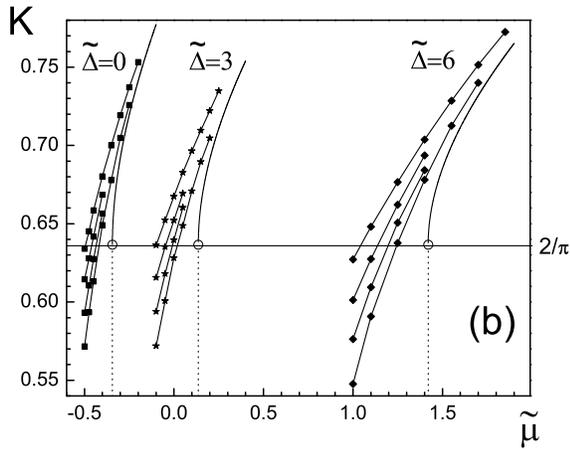}
\caption{\label{fig:FT:rho}Reduced superfluid density $K=\hbar^2\sigma_\mathrm{SF}/m^2k_\mathrm{B}T$ as a function of chemical potential $\tilde{\mu}$. Three sets of curves represent the case of no disorder $\tilde{\Delta}=0$, and disorder strengths $\tilde{\Delta}=3$ and $\tilde{\Delta}=6$. In each set, four curves from the bottom to the top correspond to finite system sizes $192\times 192$, $96\times 96$, $48\times 48$ and $24\times 24$. Statistical errors are smaller than the symbol size. (a) The upper graph shows $K$ in the wide range of $\tilde{\mu}$, and (b) the lower graph depicts proximity to the KT transition, where the solid curves are the KT extrapolation to the infinite system size. Horizontal lines are equal to the value of the universal jump $K_\mathrm{cr}=2/\pi$ at the KT transition. The dotted curve on the upper graph (a) is an example of the extrapolation of the data to lager system size $10^4 \times 10^4$.}
\end{figure}

%
%
We model systems of finite spatial size $N\times N$ to be $24\times 24$, $48\times 48$, $96\times 96$ and $192\times 192$. Three situations are considered: no disorder $\tilde{\Delta}=0$, and disorder of strengths $\tilde{\Delta}=3$ and $\tilde{\Delta}=6$. In the later cases, we average data over about 500 realizations of disorder. Temperature is kept fixed throughout all simulations, and we study how the average boson number per site $n$ and the superfluid density $\sigma_\mathrm{SF}$ behaves as we pass over the transition point increasing chemical potential $\tilde{\mu}$, see Fig.~\ref{fig:FT:n} and \ref{fig:FT:rho}. Note that it is more informative to study the dimensionless reduced superfluid density $K$ rather than $\sigma_\mathrm{SF}$ itself. According to (\ref{eqn:FT:rho}), the coefficient of proportionality relating $K$ and $\sigma_\mathrm{SF}$ is $\hbar^2/\tilde{m}^2k_\mathrm{B}T$. Thus, the universal discontinuity of the superfluid density (\ref{eqn:FT:rhoKT}) at the fixed temperature KT transition is equivalent to the jump of the reduced superfluid density $K$ from 0 to $K_\mathrm{cr}=2/\pi$.

\begin{figure}
\centering
\includegraphics[width=\columnwidth,keepaspectratio]{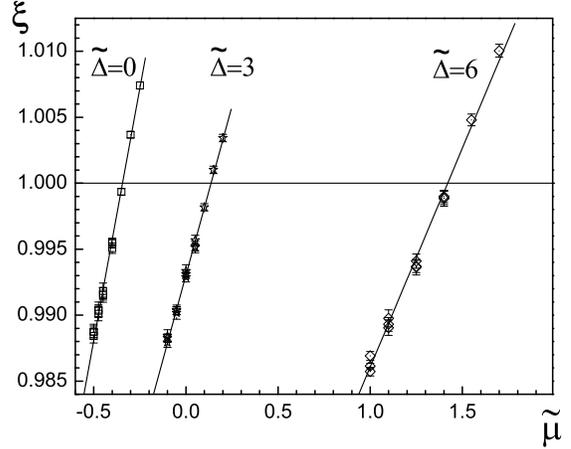}
\caption{\label{fig:FT:xi}RG parameter $\xi$ as a function of $\tilde{\mu}$. Three sets represent the cases of no disorder $\tilde{\Delta}=0$, and disorder strengths $\tilde{\Delta}=3$ and $\tilde{\Delta}=6$. In each set, values of $\xi$ are deduced by RG analysis (\ref{eqn:FT:RG}) of different pairs of $K$ corresponding to pairs of systems $192\times 192$ and $48\times 48$, $96\times 96$ and $24\times 24$, $192\times 192$ and $24\times 24$ at Fig.~\ref{fig:FT:rho}b. The solid lines are linear fits for all data points at a given disorder strength.}
\end{figure}

For a given disorder strength the absence of saturation of the reduced superfluid density $K(\tilde{\mu},N)$ for different system sizes below the critical value $K_\mathrm{cr}$ and collapse of the data above $K_\mathrm{cr}$, see Fig.~\ref{fig:FT:rho}, is a qualitative indication of the KT discontinuity at infinite system sizes. To perform quantitative analysis we use the integral form of the KT RG equations (\ref{eqn:FT:KT_K},\ref{eqn:FT:KT_y2}) (see Ref.~\onlinecite{b:exp:SFN:BoseGasPro02})
\begin{equation}
\int_{K(\tilde{\mu},N_1)/K_\mathrm{cr}}^{K(\tilde{\mu},N_2)/K_\mathrm{cr}} \frac{dt}{t^{2}(\ln t-\xi(\tilde{\mu}))+t}=-4\ln\left(\frac{N_2}{N_1}\right).\label{eqn:FT:RG}
\end{equation}
Here $\xi(\tilde{\mu})$ is the size-independent microscopic parameter characterizing the vortex fugacity, and, in contrast to the thermodynamical reduced superfluid density $K(\tilde{\mu})=K(\tilde{\mu},N\rightarrow\infty)$, $\xi(\tilde{\mu})$ is an analytic function of $\tilde{\mu}$ (i.e., $\xi(\tilde{\mu})$ may be expanded into the Taylor series $\xi(\tilde{\mu})\approx 1+\xi^\prime(0)\cdot(\tilde{\mu}-\tilde{\mu}_\mathrm{cr})+\ldots$ in the vicinity of the critical point $\mu_\mathrm{cr}$). Those two properties of $\xi$ we use to check the consistency of our data with the KT scenario of the transition. For different pairs of $K(\tilde{\mu},N)$ we solve (\ref{eqn:FT:RG}) for $\xi(\tilde{\mu})$ and observe that its values collapse on the line near $\tilde{\mu}_\mathrm{cr}$, see Fig.~\ref{fig:FT:xi}. Equation (\ref{eqn:FT:RG}) indirectly represents KT RG flow of $K(\tilde{\mu},N)$ for increasing length scales $N_1<N_2<N_3<N_4<\ldots$. In the limit $N\rightarrow\infty$ it gives equation%
\begin{equation}
\frac{K_\mathrm{cr}}{K(\tilde{\mu})}+\ln\left(\frac{K(\tilde{\mu})}{K_\mathrm{cr}}\right)=\xi(\tilde{\mu})\label{eqn:FT:RGcr}
\end{equation}
for the thermodynamical value of $K(\tilde{\mu})$. Finally, using determined function $\xi(\tilde{\mu})$, we solve (\ref{eqn:FT:RGcr}) for $K(\tilde{\mu})$, see Fig.~\ref{fig:FT:rho}b (solid curves). Note that, due to the logarithmic form of the finite-size corrections to $K$ near the KT transition,\cite{b:KT:AHNS} $K(N)-K_\mathrm{cr}\propto 1/\ln N$ for $\tilde{\mu}\gtrsim\tilde{\mu}_\mathrm{cr}$, data for systems of exponentially different sizes is required for the KT RG analysis.

The critical observation, see Fig.~\ref{fig:FT:rho}a, is that disorder dramatically increases finite-size smearing of the KT discontinuity of the superfluid density $\sigma_\mathrm{SF}\propto K$. For example, consider system of size $24\times 24$. Compared to the ideal case (disorder strength $\tilde{\Delta} =0$), there is no clearly visible characteristic bending in the data trend for the reduced superfluid density $K$ at a strong disorder ($\tilde{\Delta} =6$) near the anticipated discontinuous KT transition. And, yet, the self-consistent RG analysis of the data for exponentially different system sizes will and is revealing the KT transition.

%
%
%
%
\section{Comparison with experiments\label{s:FT:Comparison}}
%
%
%
%
Our simulations are directly applicable to explain the role of disorder in the superfluid transition in $^4$He film absorbed to microscopically rough surfaces, see Ref.~\onlinecite{b:exp:SFN:FiniteSize:Dwight}. In these experiments, sample surfaces are prepared by coating QCMs with CaF$_2$. The height distribution of a CaF$_2$ film is close to a Gaussian function, and its dispersion increases with the nominal deposition thickness (the system of zero disorder is an uncoated plain QCM crystal with typical gold electrodes). The crystals are mounted in a cell. The QCM crystals are operated in the third harmonic $f_0\approx 15\:\mathrm{MHz}$ of their fundamental resonance frequency. At the constant temperature $1.672\:\mathrm{K}$, $^4$He is incrementally added to the cell, correspondingly, the thickness of the $^4$He film absorbed to the sample surface grows. The resonance frequency of oscillations $f=\omega/2\pi$ of the QCM crystal with an absorbed $^4$He film is directly related to the viscously clamped normal fraction of the $^4$He film. When the film undergoes a transition from the normal to superfluid state, one expects to see a sharp step in the resonance frequency $f$ because of the decoupling of the zero-viscosity superfluid fraction of the $^4$He film from the substrate.\cite{b:exp:SFN:He4Chester} The absorbed film shifts the resonance frequency from the value $f_0$ by $-\Delta f$ which is proportional to the difference of total and superfluid areal densities of $^4$He, $-\Delta f\propto\sigma -\sigma_\mathrm{SF}$ (see insert of Fig.~\ref{fig:FT:DeltaF}). At a finite (angular) frequency $\omega$, one probes length scales of order of a diffusion length $r_\omega=\sqrt{14D/\omega}$, where $\omega\sim\omega _0=2\pi f_0$, and $D$ is diffusivity constant.\cite{b:KT:AHNS} This sets finite-size scaling for $\sigma_\mathrm{SF}$ and smooths out the zero frequency sharp step in $\Delta f$. It was found in Ref.~\onlinecite{b:exp:SFN:FiniteSize:Dwight} that with higher disorder no indication of this step is observable. Another signature of the superfluid transition in $^4$He films is a peak in the dissipation, and it indirectly manifested the presence of the KT transition in these experiments.
%
%
\subsection{Important length scales}
%
%
The superfluid transition in 2D bosonic systems is driven by unbinding of thermally activated vortex-antivortex pairs. Kosterlitz and Thouless used the renormalization-group technique to consider the attenuating effect of smaller pairs on the interaction between the respective members of larger pairs. They derived recursion relations for a scale-dependent superfluid density $\sigma_\mathrm{SF}(l)$ (or, corresponding reduced superfluid density $K(l)=\sigma_\mathrm{SF}(l)\hbar^2/m^2k_\mathrm{B}T$) and vortex-pair excitation probability $y^2(l)$
\begin{subequations}
\label{allequations}
\begin{eqnarray}
dK^{-1}(l)/dl &=& 4\pi^3 y^2 (l) + O[y^4 (l)],\label{eqn:FT:KT_K}\\%
dy^2 (l)/dl &=& 2[2-\pi K(l)] y^2 (l) + O[y^4 (l)],\label{eqn:FT:KT_y2}%
\end{eqnarray}
\end{subequations}
where $l=\ln{(r/a_0)}$, and contribution of vortex-antivortex pairs of sizes less than $r$ has been already taken into account. The vortex core size $a_0$ is of the order of the healing length of the superfluid, i.e., the length over which the superfluid density can change from zero to its mean value. Here, we explicitly note that KT RG equations are only valid for $y^2 \ll 1$ (for $y^2 \gg 1$ the vortex density is so high that it is ambiguous to group them into some hierarchy of vortex-antivortex pairs of different sizes and then to carry out renormalization).

For an idealized experiment carried out at zero frequency on infinite size system, the recursion relations are iterated from $a_0$ to infinite scale. In the normal phase $y^2$ diverges at infinite scales, and $\sigma_\mathrm{SF}$ is renormolized to zero. In contrast, the superfluid phase is characterized by vanishing $y^2$ and some finite $\sigma_\mathrm{SF}$. Phase transition from the superfluid to the normal phase corresponds to unbinding of vortex-antivortex pairs as they get excited to infinitely large scales of their mutual separation, and, in addition, the superfluid transition is accompanied by the universal jump (\ref{eqn:FT:rhoKT}) of ${\sigma_\mathrm{SF}}_\mathrm{cr}$.

In finite systems the maximum vortex separation is bounded by the system size, and the KT iterative renormalization process is cut off at a corresponding scale. This broadens the sharp KT transition and leads to a finite-size rounding of the KT jump in $\sigma_\mathrm{SF}$, see Fig.~\ref{fig:FT:rho}. Interpretation of experiments at finite frequencies requires the theory of Ref.~\onlinecite{b:KT:AHNS} that incorporates the dynamic response of the vortex plasma to an externally applied oscillating field. It was shown that vortex pairs of size larger than the vortex diffusion length $r_\omega=\sqrt{14D/\omega}$ cannot equilibrate to the oscillating field of the frequency $\omega$, and do not contribute to renormalization, so the iterations are truncated at $r_\omega$. Thus, results of our simulations for $\sigma_\mathrm{SF}$ for finite-size systems are in one-to-one correspondence to finite-frequency experiments, provided that $\omega$ is adjusted so that $r_\omega$ is equal to the simulated system size.

Let's specify particular values of length scales for $^4$He films absorbed to rough CaF$_2$ surfaces studied in Ref.~\onlinecite{b:exp:SFN:FiniteSize:Dwight}. Average helium film thickness in the vicinity of the superfluid transition in those experiments was reported to be about $4\:\mathrm{layers}\sim 15\:\mathrm{\AA}$ on the flat substrate (on the disordered CaF$_2$ substrate the amount of helium absorbed was larger). Temperature was in the range $1-2\:\mathrm{K}$, thus the vortex core size is expected to be on the order of atomic dimensions, $a_0\sim 10\:\mathrm{\AA}$. Because $a_0$ is comparable to the helium film thickness no 3D related vortex effects are assumed to exist. The diffusivity constant $D$ was not measured and might be affected by disorder, but from dimensional analysis $D\approx\hbar/m\sim 2\times 10^{-4}\:\mathrm{cm}^2/\mathrm{sec}$.\cite{b:KT:AHNS} (A cautionary remark: for different substrates the numerical coefficient before $\hbar/m$ in the formula for $D$ was reported\cite{b:KT:Donnelly} to be in the wide range from $0.1$ to $20$.) Dynamical measurements of $\sigma_\mathrm{SF}$ were carried out in the course of operation of the QCM crystals at $\omega\approx 2\pi\times 15\:\mathrm{MHz}$, and this gives $r_\omega\sim 500\:\mathrm{\AA}$. QCMs were of centimeter dimensions, thus their width is an irrelevant length scale. CaF$_2$ surfaces coating the QCMs consisted of peak-type structures with an average separation $R_\mathrm{s}\lesssim 100\:\mathrm{\AA}$. Because the roughness scale is less than the diffusivity length, $R_\mathrm{s}\ll r_\omega$, disorder is considered to be microscopic. In Ref.~\onlinecite{b:exp:SFN:FiniteSize:Dwight}, by varying deposition time, surfaces with average height of the CaF$_2$ structures from 0 to $\sim 50\:\mathrm{\AA}$ were prepared and used (0 stands for uncoated (plain) QCM crystals).

With respect to the lattice model (\ref{eqn:FT:Hubbard}) used in our simulations in Sec.~\ref{s:FT:Simulations}, the vortex core size $a_0$ stands for the lattice spacing $a$, and the ratio $r_\omega/a_0\sim 50$ defines lattice spatial size to be $N\times N\sim 50\times 50$. We consider height variations of the CaF$_2$ profile to act as a random potential. We chose on-site potential disorder to represent the closely spaced, raised structures of CaF$_2$. Particular choice of the disorder representation cannot affect universal properties of the second order KT transition. However, to be in line with experiments of Ref.~\onlinecite{b:exp:SFN:FiniteSize:Dwight} we must satisfy criterion of the microscopic disorder $r_\omega\gg R_\mathrm{s}$. And indeed this is true for our lattice model, because it corresponds to $R_\mathrm{s}=a\sim a_0$ and system size $r_\omega/a_0\sim N\gg 1$.
%
%
\subsection{Broadening of the KT transitions\label{s:FT:Comparison:broadening}}
\begin{figure}
\centering
\includegraphics[width=\columnwidth,keepaspectratio]{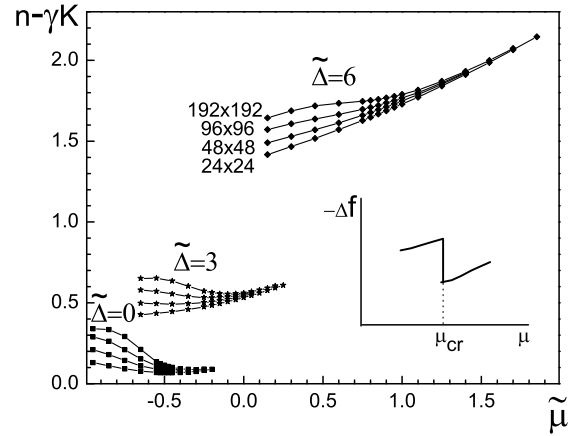}
\caption{\label{fig:FT:DeltaF}Frequency shift theoretically defined $-\Delta f\propto n-\gamma K$ as a function of chemical potential $\tilde{\mu}$. The three sets of curves represent the cases of no disorder $\tilde{\Delta}=0$, and disorder strengths $\tilde{\Delta}=3$ and $\tilde{\Delta}=6$. In each set, the four curves from the bottom to the top corresponds to finite system sizes $24\times 24$, $48\times 48$, $96\times 96$ and $192\times 192$. Statistical errors are smaller than the symbol size. The insert shows an idealized sketch of $-\Delta f(\mu)$ in the vicinity of the superfluid transition at $\mu_\mathrm{cr}$; the step corresponds to the jump of superfluid density $\sigma_\mathrm{SF}$ in the system of infinite size, see (\ref{eqn:FT:rhoKT}).}
\end{figure}

%
%
The experiments of Ref.~\onlinecite{b:exp:SFN:FiniteSize:Dwight} with $^4$He films absorbed to rough CaF$_2$ surfaces can be compared to our model in the following way (see (\ref{eqn:FT:n},\ref{eqn:FT:rho}))
\begin{equation}
-\Delta f\propto\sigma -\sigma_\mathrm{SF}\propto n-\gamma K,%
\qquad\gamma=\tilde{m}a^2k_\mathrm{B}T/\hbar^2.\label{eqn:FT:Deltaf}
\end{equation}
For our choice of parameters $\gamma\sim 1$ (we choose $\gamma =0.75$). Our simulations, specifically Fig.~\ref{fig:FT:DeltaF}, qualitatively resemble the experimental situation. Actually, Fig.~\ref{fig:FT:DeltaF} contains excessive information relative to those experiments of Ref.~\onlinecite{b:exp:SFN:FiniteSize:Dwight}. There, QCM measurements were carried out at a fixed frequency, i.e., $r_\omega$ was fixed, while CaF$_2$ substrates of different roughness were used. In other words, at a given disorder strength (i.e., on a given coated QCM) only a single curve for mass decoupling (\ref{eqn:FT:Deltaf}) as a function of the chemical potential was observed, for example, let's say that these were upper curves in each set on Fig.~\ref{fig:FT:DeltaF} for which $r_\omega$ corresponded to $192\times 192$ lattice sites in our simulations. In full agreement with experimental observations, we see that, as disorder strength is increased, finite-size behavior completely masks the step in $\Delta f$ related to the KT universal jump of the superfluid density $\sigma_\mathrm{SF}$. This is exactly the observation that did not give firm evidence to unambiguously claim the KT mechanism of the superfluid transition in Ref.~\onlinecite{b:exp:SFN:FiniteSize:Dwight}. If, at each disorder strength, QCM measurements were carried out at a series of frequencies, then similar to our results, Fig.~\ref{fig:FT:DeltaF}, one could have had enough data to perform the KT RG analysis. For further discussion see Sec.~\ref{s:FT:Discussion}.

\begin{figure}
\centering
\includegraphics[width=\columnwidth,keepaspectratio]{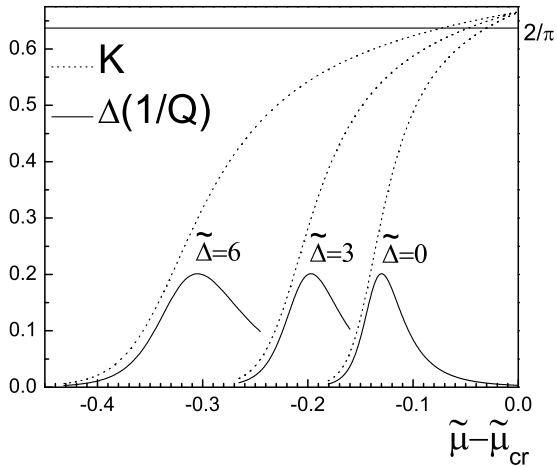}
\caption{\label{fig:FT:dissipation}Dotted lines are the reduced superfluid density $K\propto\sigma_\mathrm{SF}$ and solid lines are the dissipation peak $\Delta (1/Q)$ as a function of chemical potential $\tilde{\mu}$ for the systems of size $10^4\times 10^4$. The three pairs of curves represent the cases of no disorder $\tilde{\Delta}=0$, and disorder strengths $\tilde{\Delta}=3$ and $\tilde{\Delta}=6$. Chemical potential is shifted by the critical value $\tilde{\mu}_\mathrm{cr}$ of the KT transitions in the corresponding systems.}
\end{figure}

An essential condition of the KT transition is an observation of the dissipation peak during dynamic finite frequency measurements. It signifies change in the dynamical response from the one determined by the free vortices moving diffusively at the normal side to the one produced by the vortex pairs of size greater $r_\omega$ failing to equilibrate to external oscillating field at the superfluid side. In Ref.~\onlinecite{b:exp:SFN:FiniteSize:Dwight} even for the strongly disordered samples the peak in dissipation was present and was the basis to claim the KT scenario of the superfluid transition. Our simulations are statistical mechanical and do not simulate \emph{per se} the dissipation peak of dynamical origin. However, using the results of the linear dynamical theory for real and imaginary parts of effective dielectric constant of vortex plasma,\cite{b:KT:AHNS} we can predict the shape of the dissipation peak from our numerical data. The dissipation is measured in terms of the change in quality factor $Q$ that is given by\cite{b:He:Mylar:AgnoletExp,b:KT:Donnelly}%
\begin{equation}
\Delta \left(\frac 1Q\right)\propto%
\frac{\pi^2}{2}\cdot\frac{(Ky)^2}{1+(\pi^4Ky^2)^2}.\label{eqn:FT:Q}
\end{equation}%
The dissipation peak is positioned close to the point of the inflection of the growing $\sigma_\mathrm{SF}$ as one goes from the normal to the superfluid phase while increasing the control parameter, chemical potential $\mu$. But KT RG works in a narrow window of the transition, e.g., we had to use only a portion of the simulated superfluid density data Fig.~\ref{fig:FT:rho}b to deduce from (\ref{eqn:FT:RG}) the size-independent microscopic parameter $\xi$. Too deep in the normal phase $y^2\gtrsim 1$, and the KT recursion relations (\ref{eqn:FT:KT_K},\ref{eqn:FT:KT_y2}) break down. Thus, we had to extrapolate our data to higher system sizes, so that $y^2\lesssim 1$ and the dissipation peak is positioned in the same range of $\tilde{\mu}$. For each disorder strength, we used linear fits for $\xi(\tilde{\mu})$ from Fig.~\ref{fig:FT:xi} to recalculate the finite-size smeared reduced superfluid density $K(\tilde{\mu},N=192)$ of the system of size $192\times 192$ into $K(\tilde{\mu},N=10^4)$.\cite{f3} The value of $y^2(\tilde{\mu},N)$ for known $K(\tilde{\mu},N)$ and $\xi(\tilde{\mu})$ is given by%
\begin{equation}
y^2=\frac{1}{2\pi^2}\left[\frac{K_\mathrm{cr}}{K}+\ln\left(\frac{K}{K_\mathrm{cr}}\right) -\xi\right].\label{eqn:FT:y2}
\end{equation}%
Our observations for $\Delta(1/Q)$, see Fig.~\ref{fig:FT:dissipation}, are that for one and the same $r_\omega$ with an increase of disorder strength the dissipation peak will be broadened and will be shifted deeper in the normal phase away from the critical point of the superfluid transition. A high level of noise in the data for $^4$He films coating rough CaF$_2$ surfaces, Ref.~\onlinecite{b:exp:SFN:FiniteSize:Dwight}, did not allow a validation of this statement (note that dissipation peaks were clearly seen for all values of disorder). However, in those experiments there is room left to raise accuracy at least by the order of magnitude.\cite{b:exp:SFN:FiniteSize:DwightPrivate}


We have also studied finite-size scaling of the width of the dissipation peak for a given disorder strength. Its broadening one-to-one resembles the finite-size broadening of the superfluid density jump (correspondingly, position of the peak shifts to higher chemical potential as $r_\omega$ is increased).
%
%
\subsection{Disorder description in detail\label{s:FT:Discussion:Disorder}}
%
%
Our simulations explain broadening of the KT transition in \emph{microscopically} disordered 2D bosonic systems such as the $^4$He films absorbed to rough CaF$_2$ substrates in Ref.~\onlinecite{b:exp:SFN:FiniteSize:Dwight}. In our case, $R_\mathrm{s}\ll r_\omega$, the roughness scale $R_\mathrm{s}$ is much less compared with the dimension $r_\omega$ of the region effectively studied at finite frequency. Different case of \emph{macroscopic} disorder, $R_\mathrm{s}\gg r_\omega$, can be achieved, for example, in experiments with $^4$He films absorbed on a Mylar substrate when the torsional oscillator technique is implemented.\cite{b:He:Mylar:AgnoletExp,b:He:Mylar:AgnoletTheory} In these cases, typical $\omega\approx 2\pi\times 1500\:\mathrm{Hz}$ leads to $r_\omega\sim 5\:\mu\mathrm{m}$. Mylar contains bubbles and flakes of sizes $1-100\:\mu\mathrm{m}$ and average separation of $10-150\:\mu\mathrm{m}$, i.e., $R_\mathrm{s}\sim 100\:\mu\mathrm{m}$. Macroscopic variations of the substrate potential result in the inhomogeneous broadening of the KT type superfluid transition. Similar to our suggestion, frequency dependence of the inhomogeneous broadening of the jump of the superfluid density and the dissipation peak was successfully studied by a two-torsional-oscillators technique.\cite{b:He:Mylar:torsialExp}

The fact that in our simulations we relate rough surface of the CaF$_2$ substrate of Ref.~\onlinecite{b:exp:SFN:FiniteSize:Dwight} with the spatially varying potential implies that helium film thickness is modulated by the microscopic surface roughness. This is true for the case of the superfluid transition at high temperatures (in Ref.~\onlinecite{b:exp:SFN:FiniteSize:Dwight} $T=1.672\:\mathrm{K}$) or, equivalently, at high average helium film thicknesses, see the phase diagram Fig.~\ref{fig:FT:phaseDIAG}. For $T\gtrsim 1\:\mathrm{K}$ we expect that the surface tension predominates, and microscopic puddles of helium are created in the valleys of the disordered surface profile while the helium film thins close to the ridges of the peak-type structures of the CaF$_2$ substrate. A study similar to Ref.~\onlinecite{b:exp:SFN:FiniteSize:Dwight} of the superfluid transition of $^4$He films absorbed to the same rough CaF$_2$ substrates was carried out at low temperatures $T\lesssim 1\:\mathrm{K}$. It turned out that in this case broadening of the KT transition was almost independent of the strength of disorder.\cite{b:exp:SFN:FiniteSize:DwightPHD} This allows us to speculate that a different regime from the one considered in this paper was achieved for $T\lesssim 1\:\mathrm{K}$, when van der Waals' forces win over the surface tension and the helium film is uniformly flattened over the rough surface, and, hence, the surface potential is uniform.
%
%
%
%
\section{Related Physical Systems\label{s:FT:Related}}
%
%
%
%
Our research is centered around the KT-type 2D finite temperature superfluid transitions in $^4$He films. Phase transitions in thin-film superconductors,\cite{b:exp:SFN:Mooij} Josephson junction arrays,\cite{b:exp:SFN:JJA} weakly interacting 2D Bose gases,\cite{b:exp:SFN:BoseGasPro02,b:exp:SFN:BoseGasPro01} 2D gas of spin-polorized atomic hydrogen on liquid-helium surface,\cite{b:exp:SFN:H} and Bose-Enstein condensates loaded on a 2D optical lattice\cite{b:exp:SFN:TrombettoniNewJ} belong to the same universality class. Our analysis of the role of diagonal disorder in $^4$He films can be qualitatively extended to some of these cases.

Microscopic details are irrelevant for the study of universal properties of the second order KT phase transition characterized by the diverging correlation length scales.\cite{b:SFI:FWGF89,b:WAsim:MAIN:WSGY} For example, universal jump (\ref{eqn:FT:rhoKT}) of the areal superfluid density is innate for both discrete lattice type and continuous geometry of experimental realizations of 2D bosonic systems with diagonal disorder and controlled value of the chemical potential. Instead of $^4$He film absorbed to rough surface with the chemical potential controlled by the pressure in the sample cell,\cite{b:exp:SFN:FiniteSize:Dwight} we could have considered as well an atomic gas in an optical lattice with a given number of atoms per site and a varying trapping potential, or a Josephson junction array (or a superconducting granular film) with an external voltage applied to the ground plane and random gate voltages induced by trapped charge in the substrate.\cite{b:SFI:exp:vanderZantcJJA} Critical behavior in all mentioned cases falls within the framework of the lattice bosonic Hubbard model (\ref{eqn:FT:Hubbard}). The same finite-size analysis as in Sec.~\ref{s:FT:Simulations:results} can be used to resolve the universal KT jump (\ref{eqn:FT:rhoKT}). However, microscopic details determine how strong will be finite-size effects. Thus, similar to our study targeting $^4$He films absorbed to rough surfaces in Ref.~\onlinecite{b:exp:SFN:FiniteSize:Dwight}, separate simulations are required for the particular experimental situation to reveal (or disproof) primary importance of finite-size effects as a source of experimental smearing of the discontinuous KT transition.

Along with diagonal disorder, off-diagonal disorder can take place in 2D systems that exhibit finite temperature KT transition. Off-diagonal disorder can be present in the form of site or bond dilution disorder in Josephson junction arrays \cite{b:PT:d:HarrisJJA}, or as nonmagnetic disorder in planar symmetry spin models \cite{b:PT:d:brazil}, or random gauge disorder \cite{b:PT:d:Scheidl1} (such as disorder in the phase of the complex-number tunnelling amplitudes for bosonic systems), etc. Our study of non-universal corrections to the finite-size broadening of the KT transition produced by diagonal disorder cannot be directly related to the case of off-diagonal disorder. 2D $XY$ models are used for description of aforementioned disordered cases. But an $XY$ model simply does not admit conventional diagonal disorder. Formally within the bosonic Hubbard hamiltonian (\ref{eqn:FT:Hubbard}) the $XY$ limit is obtained by writing the last two terms as $U/2\left(\hat{n}_i-(\mu-v_i)/U\right)^2$ and setting simultaneously $U\rightarrow\infty$, and $\mu\rightarrow\infty$. In this limit average number of particles per site is fixed, $n=\mu/U$, and there are no local density fluctuations at all, i.e., any finite diagonal disorder $v_i$ in $\mu$ is ignored with respect to the $XY$ models.

On a passing note, finite-size effects and disorder will play an important role in the observation of the superconducting transition in 2D (the KT scenario is debated now for low-$T_c$ granular films,\cite{b:SC:FiniteSize:Lemberger1} high-$T_c$ films,\cite{b:SC:FiniteSize:Lobb1} and Josephson junction arrays.\cite{b:SC:FiniteSize:Lobb2}) However, we do not continue our discussion of both neutral superfluid and superconducting [charged superfluid] systems due to differences in the corresponding measurement techniques, and due to additional magnetic field screening specific for superconducting systems.\cite{b:SC:FiniteSize:Pierson}
%
%
%
%
\section{Discussion and conclusions\label{s:FT:Discussion}}
%
%
%
%
We have studied a 2D bosonic system with diagonal disorder in the vicinity of the finite temperature superfluid transition. We found that the disorder dramatically increases finite-size smearing of the universal jump of the superfluid density (\ref{eqn:FT:rhoKT}). And, an analysis of its finite-size scaling is required to resolve faithfully the underlying KT scenario of the superfluid transition.

Our simulations consistently explained difficulties of identification of the KT transition by QCM frequency shift in the experiments of Ref.~\onlinecite{b:exp:SFN:FiniteSize:Dwight} with $^4$He absorbed to microscopically rough CaF$_2$ surfaces. In those experiments QCM measurements were carried out at a \emph{fixed} frequency $\omega$. Thus, for a given disorder strength, the effective finite length scale set by the vortex diffusivity length $r_\omega\propto 1/\sqrt{\omega}$ was fixed. Without finite-size scaling it was not possible to extract the universal jump of the superfluid density (\ref{eqn:FT:rhoKT}) in the case of strongly disordered substrates. We speculate that operation of QCM crystals at \emph{different} harmonics of their fundamental resonance will be required. Note that QCM measurements of the superfluid transition of $^4$He films on a flat substrates at two different harmonics of fundamental resonance frequency were carried out. Finite-size broadening that increased at higher frequencies was observed.\cite{b:exp:SFN:FiniteSize:Glaberson93} Another study of the $\omega$ dependance of the KT transition in 2D $^4$He films used an ultrasonic technique.\cite{b:exp:SFN:FiniteSize:ultrasound} For a given sample the superfluid transition was relatively broad at high frequencies with the universal jump of the superfluid density smeared toward the normal phase, but sharpened as the frequency was decreased (correspondingly, the position of the dissipation peak was shifting). Our simulations (Sec.~\ref{s:FT:Comparison:broadening}) predict the same scenario for a given disorder strength as $r_\omega$ is increased.

The experiments of Ref.~\onlinecite{b:exp:SFN:FiniteSize:Dwight} were our original motivation. Due to the noise in the measured data we sought only a qualitative description of those experiments. Thus, we used a lattice Hubbard model with on-site potential disorder to describe helium films and discrete space and imaginary time worm algorithm to carry out classical Monte Carlo simulations (Sec.~\ref{s:FT:Simulations}). Universal properties of the superfluid transition were preserved, but one-to-one matching of the experiments and the simulations cannot be achieved in this kind of simulations setup. Exact \emph{ab initio} numerical simulations of $^4$He film with realistic substrate potentials and geometry is possible with extension of the used worm algorithm\cite{b:WA:ProSvi1} for continuous-space path Monte Carlo simulations.\cite{b:WA:ProSvi2} Another aspect is that, with regard to the experiments of Ref.~\onlinecite{b:exp:SFN:FiniteSize:Dwight} on $^4$He films, we have considered the case of a 2D bosonic system driven to the point of the superfluid transition by the increase of boson density (i.e., chemical potential was the control parameter). We expect qualitatively similar results for broadening of the KT transition if, instead, at a constant boson density the system will be brought across the superfluid--normal phase boundary by an increase of the temperature (see Fig.~\ref{fig:FT:phaseDIAG}).
%
%
%
\begin{acknowledgments}
Author is very grateful R.B. Hallock and D.R. Luhman for illuminating discussions and for explanation of details of experiments carried out with $^4$He films absorbed to rough surfaces. Thanks are also due to N.~Prokofev and B.~Svistunov for a large number of valuable suggestions and discussions. The research was supported by the National Science Foundation under Grant No. PHY-0426881, and, in part, under Grant No. PHY99-07949.
\end{acknowledgments}
%
%
%
%

%

\begin{thebibliography}{999}
\bibitem{f3} Equation (\ref{eqn:FT:RG}) was used along with some polynomial fits of the data in Fig.~\ref{fig:FT:rho}b for $K$ for the system of size $192\times 192$ for each disorder strength. The range of $\tilde{\mu}$ for $K(\tilde{\mu},N=10^4)$ in Fig.~\ref{fig:FT:dissipation} corresponds to the ones in Fig.~\ref{fig:FT:rho}b and Fig.~\ref{fig:FT:xi}.
\end{thebibliography}
\end{document}